\documentclass[reprint,prb,showkeys,
superscriptaddress,citeautoscript] {revtex4-1}

\usepackage{graphicx}
\usepackage{tikz}
\usepackage{braket}
\usepackage{hyperref}
\hypersetup{colorlinks=true, urlcolor= blue, citecolor=blue, linkcolor= blue, 
bookmarks=true, bookmarksopen=false}

\usepackage{amsmath,amssymb}
\usepackage{textcomp}
\usepackage{multirow}

\usepackage{enumitem}
\setlist{noitemsep, leftmargin=*}

\usepackage{times}
\begin{document}
\title{Electronic and Excitonic Properties of Semi-Hydrogenated Borophene Sheets}
\author{Mohammad Ali Mohebpour}

\author{Shobair Mohammadi Mozvashi}
\affiliation{Computational Nanophysics Laboratory 
	(CNL), Department of Physics, University of Guilan, P. O. Box 41335-1914, 
	Rasht, Iran.}

\author{Sahar Izadi Vishkayi}
\affiliation{School of Physics, Institute for Research in Fundamental 
	Sciences (IPM), P. O. Box 19395-5531, Tehran, Iran.}

\author{Meysam Bagheri Tagani}
\email{m{\_}bagheri@guilan.ac.ir}
\affiliation{Computational Nanophysics Laboratory 
	(CNL), Department of Physics, University of Guilan, P. O. Box 41335-1914, 
	Rasht, Iran.}

\begin{abstract}

Borophene has triggered a surge of interest due to its outstanding properties 
including mechanical flexibility, 
polymorphism, and opto-electrical anisotropy. Very 
recently, a novel semi-hydrogenated borophene, called \mbox{$\alpha'$-4H}, was 
synthesized in large-scale freestanding samples, which exhibits excellent 
air-stability and semiconducting nature. Herein, using the density functional 
theory (DFT) and many-body perturbation theory (MBPT), we investigate the 
electronic and excitonic optical properties of \mbox{$\alpha'$-4H} borophene. 
The 
DFT results reveal that by breaking the mirror symmetry and increasing the 
buckling 
height 
of pure $\alpha'$ borophene, hydrogenation causes an orbital 
hybridization and opens an indirect band gap of 1.49 eV in 
$\alpha'$-4H borophene. This 
value is corrected to be 1.98, 2.23, and 2.52~eV under the G$_0$W$_0$, GW$_0$, 
and 
GW levels of theory, respectively. 
The optical spectrum achieved from solving the Bethe-Salpeter equation shows an 
optical band gap of $\sim$~2.40~eV, which corresponds to a strongly bound and 
stable bright exciton with a binding energy of $\sim$~1.18~eV. More 
importantly, the excitonic states are robust against tension up to 10\%, where 
the monolayer is dynamically stable. We also 
design and study the bilayer \mbox{$\alpha'$-4H} borophene with different 
stackings. 
For the weak 
van der Waals interactions between the layers, the bilayer can preserve most of 
the structural and electronic properties of the monolayer. Our 
study exposes the underlying physics behind the structural, electronic, and 
optical properties of \mbox{$\alpha'$-4H} borophene and suggests it as a very 
promising candidate for flexible optoelectronic applications.

\end{abstract}

%\keywords{My Keywords}
\maketitle

\section{Introduction}

Two-dimensional (2D) boron, also known as borophene, is recently considered 
as a game-changing player of future electronics because of its exceptional 
properties including polymorphism, electronic and optical anisotropy, ultrahigh 
thermal conductance, optical 
transparency, and phonon-mediated superconductivity \cite{mannix18, 
liu20-frAQB, penev16, vishkayi17, vajary18, vishkayi18}. Moreover, short B-B 
bonds and low atomic mass leads to very outstanding mechanical characteristics 
in 
borophene \cite{mannix15}. Numerous studies have studied the applications of 
borophene in rechargeable ion batteries, gas storages, catalysts, and sensors 
\cite{arabha20, ayodhya20}. Several recent studies have focused on the use of 
borophene in 
flexible electronics, wearable devices, and medicines, owing to its flexible 
nature, low mass density, and non-toxicity \cite{hou20, xie20}.

Boron has three valance electrons and, to compensate its octet completion 
deficiency, it tends to make highly-delocalized multicenter bonds, resulting in 
a vast structural diversity and polymorphism \cite{entwistle02}. Several motifs 
of borophene have been proposed and synthesized up to now, among which are 
$\alpha$, $\alpha'$, $\beta$, and $\chi$ phases, which have trigonal structures 
with 
different 
numbers 
and 
positions of hexagon holes (HHs) \cite{kong19, powar19}. Different phases of 
borophene are 
defined in terms of parameter $\nu = n/N$, where $n$ and $N$ are the numbers of 
HHs 
and 
boron atoms in a unit cell, respectively. Most of the physical properties of 
borophenes 
strongly depend on this morphology variable, thus, there is a wide set 
of options to reach a borophene phase with desired properties. Furthermore, 
unlike other 2D materials, borophene undergoes a structural phase transition, 
rather than direct fracturing under tensile strain \cite{mannix15}.

Due to its atomic arrangement, borophene cannot be exfoliated mechanically. 
Therefore, other methods such as chemical vapor deposition, molecular beam 
epitaxy, and liquid phase exfoliation should be used \cite{bhattacharyya17}. 
Moreover, because of 
its electron deficiency, borophene cannot be stable in a free-standing form and 
support of a metal substrate is required \cite{suehara10}. Several successful 
syntheses have been reported on Ag, Cu, Ni, and Au substrates \cite{wu19, 
kiraly19, zhong17}. It is 
worth mentioning that the Ag, Cu, and Ni substrates donate electrons to, and Au 
substrate withdraws electrons from borophene, hence, the borophenes grow flat 
on the former substrates and buckled on the latter one \cite{mannix15, liu13}.

Because of its high reactivity, most of boron nanostructures suffer from 
rapid oxidation and presence of other ambient contaminations \cite{mannix15}. 
Surface modification is found to be a capable tool for 
enhancing the stability and electronic properties of nanostructures 
\cite{ryder16, johns13}. 
Specifically, 
hydrogenation is reported widely to have a major impact on the 
stability of borophene and other two-dimensional materials \cite{li21, 
mohebpour20}.

Recently, Hou et al \cite{hou20} have prepared a novel semi-hydrogenated 
borophene 
phase in 
large quantities by in-situ decomposition of NaBH$_4$ under the exposure of 
hydrogen. This borophene was named \mbox{$\alpha'$-4H} 
because it 
is similar to $\alpha'$ borophene, with 4 adsorbed hydrogen atoms in each unit 
cell. The 
\mbox{$\alpha'$-4H} 
borophene has three major advantages which make it very attractive. First, it 
has high dynamical stability in freestanding form: by withdrawing 
electrons from the boron atoms, hydrogenation acts 
similar to an Au substrate, resulting a stable buckled structure. Second, it 
is stable in the air for a long time: 
the screening of hydrogen passivates the surface of borophene and prevents it 
from reactions with oxygen or other ambient atoms. Third, unlike other 
phases, 
\mbox{$\alpha'$-4H} borophene has an energy band gap of $\sim$2.5~eV, which 
brings 
hopes for its applications in optoelectronics as a semiconductor.

One of the most important drivers of recent advances in condensed matter 
physics is the feedback Ping-Pong between theory and experiment. High-quality 
theoretical and computational investigations can truly guide the experiments 
towards novel materials with desired properties. Moreover, discussing  the 
underlying physics of quantum systems improves the physical insights and 
enhances the theory for future advances. To have a small contribution to this 
trend, in our previous work, we systematically investigated the mechanical 
flexibility and strength of \mbox{$\alpha'$-4H} borophene by first-principle 
calculations 
\cite{mozvashi21}. We also demonstrated dynamical stability, mechanical 
anisotropy, high 
strain 
compliance, and tunable band gap for this monolayer.

In this paper, we investigate the electronic and excitonic optical 
properties of \mbox{$\alpha'$-4H} borophene using ab-initio many-body 
calculations. By density functional theory (DFT) calculations, we 
initially discuss the microscopic aspects in which the metallic 
$\alpha'$ 
borophene turns into the semiconducting \mbox{$\alpha'$-4H}. We show how 
hydrogenation leads to an orbital hybridization and opens a wide band gap of 
1.49~eV at the PBE level of theory.

Many-body effects (electron-electron and electron-hole 
interactions) are significantly enhanced in low-dimensional systems %including 
%quasi-1D nanowires [26], nanotubes [27], nanoribbons [28], and 2D monolayers 
\cite{cudazzo10, yang09}. This is mainly due to the reduced Coulomb screening 
and geometrical 
confinement. Hence, we expect to see the critical influence of 
electron-electron and electron-hole interactions on the electronic and optical 
properties 
of \mbox{$\alpha'$-4H} borophene.
First, we calculate the quasiparticle band gap to be 1.98, 2.23, and 2.52~eV 
at the G$_0$W$_0$, GW$_0$, and GW levels of many-body calculations, 
respectively. Then, we solve the 
Bethe-Salpeter equation on top of quasiparticle energies (the so-called 
GW+BSE) to obtain the optical spectrum. It shows that the 
direct optical excitation is characterized by an optical band gap of 
$\sim$~2.40~eV, which corresponds to a strongly bound and stable Frenkel 
exciton with a binding energy of $\sim$~1.18~eV, making \mbox{$\alpha'$-4H} 
borophene a desirable candidate for exploring optical applications. 

Finally, we design and 
investigate the 
bilayer \mbox{$\alpha'$-4H} borophene with different stackings and discuss that 
for the 
weak 
van der Waals interaction between the layers, the bilayer can preserve most of 
structural and electronic properties of the monolayer and can be used for 
similar applications. Our results bring insights about physical properties of 
the monolayer and 
bilayer \mbox{$\alpha'$-4H} 
borophene and suggest it as a promising material for flexible optoelectronic 
applications.

\section {Theoretical Background}

We start the density functional theory (DFT) calculation to find the 
ground-state eigenvalues ($E_{nk}$) and eigenfunctions ($\Phi_{nk}$) by solving 
the Kohn-Sham equation given below:
\begin{equation}
(T+V_{n-e}+V_H+V_{xc})\Phi_{nk}=E_{nk}\Phi_{nk},
\end{equation}

\noindent where T is the kinetic energy operator, $V_{n-e}$ the Coulomb 
potential of the 
nuclei, $V_H$ the Hartree potential, and $V_{xc}$ the exchange-correlation 
potential. The quasiparticle (QP) energies ($E_{nk}^{QP}$) are determined 
within 
the 
GW approximation as a first-order perturbative correction to the DFT 
single-particle energies ($E_{nk}$) \cite{shishkin07, hybertsen86}:
\begin{equation}
    \begin{aligned}
	E_{nk}^{QP,1}=Re[\bra{\Phi_{nk}}T+V_{n-e}+V_H+ 
	\Sigma_{xc}(G,W;E_{nk})\\\ket{\Phi_{nk}}] = E_{nk} + 
	\braket{\Phi_{nk}|\Sigma^{GW}-V_{xc}|\Phi_{nk}},
	\end{aligned}
	\end{equation}
	
\noindent where $\Sigma=iG_0 W_0$ is the self-energy operator. This equation is 
solved 
non-self-consistently. Therefore, it is the simplest level of theory (the 
so-called G$_0$W$_0$) and computationally the most efficient one. It calculates 
the QP energies from a single iteration by neglecting all off-diagonal elements 
of the self-energy and employing a Taylor expansion of the self-energy around 
the DFT energies. The noninteracting Green’s function ($G_0$) is directly 
calculated from single-particle orbitals and energies as:

\begin{equation}
G_0(r,r',\omega)=\sum \frac{\Phi_{nk}(r)\Phi_{nk}^*(r')}{\omega-E_{nk}-i\eta~ 
sgn(E_F-E_{nk})},
\end{equation}

\noindent where $E_F$ is the Fermi energy and $\eta$ a positive infinitesimal 
quantity. 
The noninteracting screened Coulomb potential ($W_0$) is obtained by the 
inverse dielectric function ($\varepsilon^{-1}$) and bare Coulomb interaction 
($\upsilon$) as below \cite{liu20}:

\begin{equation}
W_0(r,r',\omega')=\int dr''\varepsilon^{-1}(r,r'',\omega')\upsilon(r'',r').
\end{equation}

At higher levels of theory (i.e. G$_i$W$_0$ and G$_i$W$_i$), the updated QP 
energies ($E_{nk}^{QP,i+1}$) are achieved from the QP energies of the 
previous iteration as follows:

\begin{equation}
\begin{aligned}
E_{nk}^{QP,i+1}=E_{nk}^{QP,i}+Z_{nk}Re[\bra{\Phi_{nk}}T+V_{n-e} 
\\+V_H+\Sigma_{xc}(G,W;E_{nk})\ket{\Phi_{nk}}-E_{nk}^{QP,i}],
\end{aligned}
\end{equation}

\noindent where the renormalization factor ($Z_{nk}$) is given by 
\cite{karlicky13, shahrokhi16, shishkin06}:

\begin{equation}
Z_{nk}=(1-Re\bra{\Phi_{nk}}\frac{\partial}{\partial\omega}\Sigma(\omega)
	_{|_{E_{nk}^{QP,i}}}\ket{\Phi_{nk}})^{-1}.
\end{equation}

At the GW$_0$ level of theory, the Green’s function ($G$) is iteratively 
updated while the screened Coulomb potential ($W$) is kept fixed at the initial 
DFT value. At the GW level, both of them are reevaluated at each iteration 
($i$). 
Here, the exact (interacting) Green’s function is linked to its non-interacting 
version with the Dyson equation given as, \mbox{$G=G_0+G_0 \Sigma G$} 
\cite{blase20}.
 
The 
imaginary part of the macroscopic dielectric function is obtained by summation 
over the empty conduction bands within the long-wavelength limit ($q$ 
$\rightarrow$ 0) \cite{gajdos06}:

\begin{equation}
\begin{split}
Im\varepsilon_{\alpha\beta}(\omega)=\frac{4\pi^2e^2}{\Omega}\lim_{q \to 
0} \frac{1}{|q|^2}\sum_{c,v,k} 2w_k
\delta(\varepsilon_{ck}-\varepsilon_{v k}-\omega)\\
\times\braket{u_{ck+e_\alpha q}|u_{v k}}\braket{u_{ck+e_\beta q}|u_{v k}}^*,
\end{split}
\end{equation}

\noindent where $q$ is the Bloch vector of the incident wave, $w_k$ the 
\mbox{\mbox{$k$-point}} 
weight, $\Omega$ the volume of the unit cell, and $u_{ck}$ the cell periodic 
part of the wave function. The vectors $e_\alpha$ are unit vectors for the 
three Cartesian directions. The indices $c$ and $v$ correspond to the 
conduction and valence band states, respectively. The electron-hole excited 
states are characterized by the expansion \cite{shahrokhi17}:

\begin{equation}
\ket{S}=\sum_{c}^{elec} \sum_{v}^{hole} \sum_{k}A_{cv k}^S\ket{cv 
k},
\end{equation}

\noindent where $A^S$ is the amplitude of a free electron-hole pair 
configuration 
composed of the electron state $\ket{ck}$ and the hole state $\ket{vk}$. It is 
achieved from the diagonalization of the following excitonic equation 
\cite{rohlfing98, rohlfing00}:

\begin{equation}
\begin{split}
(E_{ck}^{QP}-E_{v k}^{QP})A_{c v k}^S + \sum_{c' v' k'}\braket{c v 
k|\Xi^{e-h}|c' v' k'}
A_{c' v' k'}^S \\= \Omega^s A^s_{c v k},
\end{split}
\end{equation}

\noindent where $E_{ck}^{QP}(E_{v k}^{QP})$ denotes the QP eigenvalues of 
valance (conduction) band at a specific \mbox{$k$-point}, kernel $\Xi^{e-h}$ 
describes 
the screened interaction between excited electrons and holes, and $\Omega^S$ is 
the excitation energy.

\section{Computational Details}

The first-principles calculations were performed in the framework of DFT, using 
the generalized gradient approximation developed by Perdew-Burke-Ernzerhof 
(GGA-PBE) and the projector augmented waves (PAW) pseudopotentials implemented 
in the Quantum ESPRESSO package \cite{giannozzi09}. The energy cutoff for the 
plane-wave 
basis set was set to be 60 Ry. The Brillouin zone was integrated with a 
\mbox{13$\times$ 13 $\times$ 1} $\Gamma$-centered \mbox{$k$-point} mesh. The 
atomic 
positions 
and 
lattice parameters were fully relaxed until a force convergence of $10^{-3}$ 
eV/\AA\ was achieved. A vacuum space of $L_z$ = 20~\AA\ was considered 
%along the $z$-direction 
to avoid spurious interactions.

The many-body perturbation calculations were carried out using the GW 
approximation. The eigenvalues and eigenstates achieved from DFT-PBE served as 
input to calculate the QP energies. The calculations were performed under 
different levels of self-consistency including single-shot (i.e. G$_0$W$_0$) 
and 
partially self-consistent (i.e. GW$_0$ and GW). A compromised set of input 
parameters (e.g. 9 $\times$ 9 $\times$ 1 \mbox{\mbox{$k$-point}} mesh, $L_z$ = 
25~\AA 
, 
40 
virtual bands, 
and 
4 
iterations) was utilized for our GW calculations. 

The QP band structure was 
interpolated at the G$_0$W$_0$ level using the maximally localized Wannier 
functions 
(MLWFs) implemented in the WANNIER90 code \cite{mostofi14}. Here, the $sp^3$ 
and $s$ 
orbitals 
of 
B and H atoms were respectively chosen for the initial projections. The 
excitonic optical properties were investigated by solving the Bethe-Salpeter 
equation (BSE) on top of GW eigenvalues and wave functions, using the 
Tamm-Dancoff approximation (TDA). This approximation takes only the resonant 
part of the BS Hamiltonian into account. The 10 highest valence bands and the 
10 lowest conduction bands were considered as the basis for the excitonic 
eigenstates.

\

\section{Results and Discussion}
\subsection{Structural and electronic properties}

\begin{figure*}
	{\includegraphics[width=0.76\textwidth]{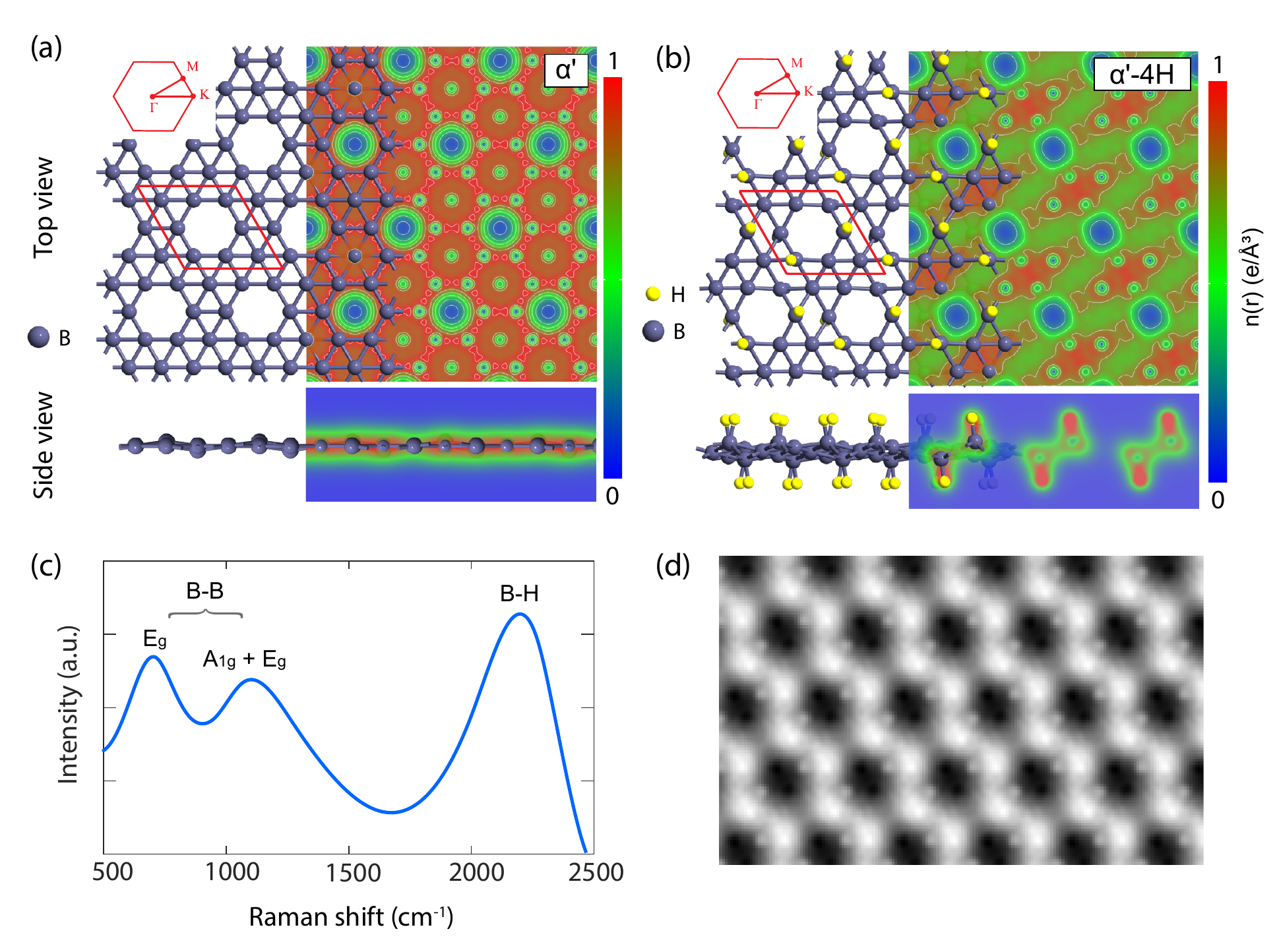}}
	\caption{\textbf{(a, b)} Structural configurations and electron density cut 
		planes of $\alpha'$ and \mbox{$\alpha'$-4H} borophenes. \textbf{(c)} 
		Simulated Raman spectrum and \textbf{(d)} simulated STM image of 
		\mbox{$\alpha'$-4H} borophene.}
	\label{fig1}
\end{figure*}

The structural configurations of fully relaxed $\alpha'$ and 
\mbox{$\alpha'$-4H} 
borophenes are displayed in FIG.~\ref{fig1} (a, b) and the corresponding 
structural 
parameters are listed in Table S1. As can be seen, the pure $\alpha'$ borophene 
is nearly flat 
($\Delta$~=~0.37~\AA) and isotropic, with a hexagon hole concentration of 
$\nu~=~1/8$. 
The hydrogenation leads to an increase in the buckling in 
\mbox{$\alpha'$-4H} 
borophene ($\Delta$~=~0.88~\AA), owing to the attraction between B and H atoms. 
The 
lattice constants of the two monolayers are similar (5.05~\AA) but the isotropy 
of the lattice breaks after hydrogenation. As a result, the bond length 
increases from isotropic 1.68~\AA\ to anisotropic 1.69 and 1.77~\AA.

FIG.~
\ref{fig1} (a, b) also 
indicate the 
electron density cut planes of  $\alpha'$ and \mbox{$\alpha'$-4H} borophenes. 
In 
$\alpha'$ borophene, we can see a “sea of electrons”, which is apparently the 
cause of its metallic nature. However, in \mbox{$\alpha'$-4H} borophene, an 
electron 
accumulation takes place around H atoms, which lowers the in-plane carrier 
distribution and decreases the electronic conductivity.

As mentioned above, the \mbox{$\alpha'$-4H} borophene was synthesized recently 
\cite{hou20}. To validate our model with the experiment, we 
performed 
simulations of Raman spectrum and scanning tunneling microscopy (STM). FIG.~
\ref{fig1} 
(c) represents the calculated Raman spectrum of \mbox{$\alpha'$-4H} borophene. 
We can 
see three peaks at about 700, 1100, and 2200 cm$^{-1}$. The first two peaks are 
representatives of $E_g$ and $A_{1g}+E_g$ modes of the B-B cluster, and the 
last peak 
is 
associated with the combined modes of the \mbox{B-H} bonds. We also indicated 
the 
simulated STM image of \mbox{$\alpha'$-4H} borophene in FIG.~\ref{fig1} (d), 
which is 
very similar to the experimentally obtained image. Our calculated Raman 
spectrum and simulated STM image support the experimental data very well 
\cite{hou20}, 
therefore, we can assure that our simulated model is completely similar to 
the synthesized \mbox{$\alpha'$-4H} borophene.

\begin{figure*}
	{\includegraphics[width=0.77\textwidth]{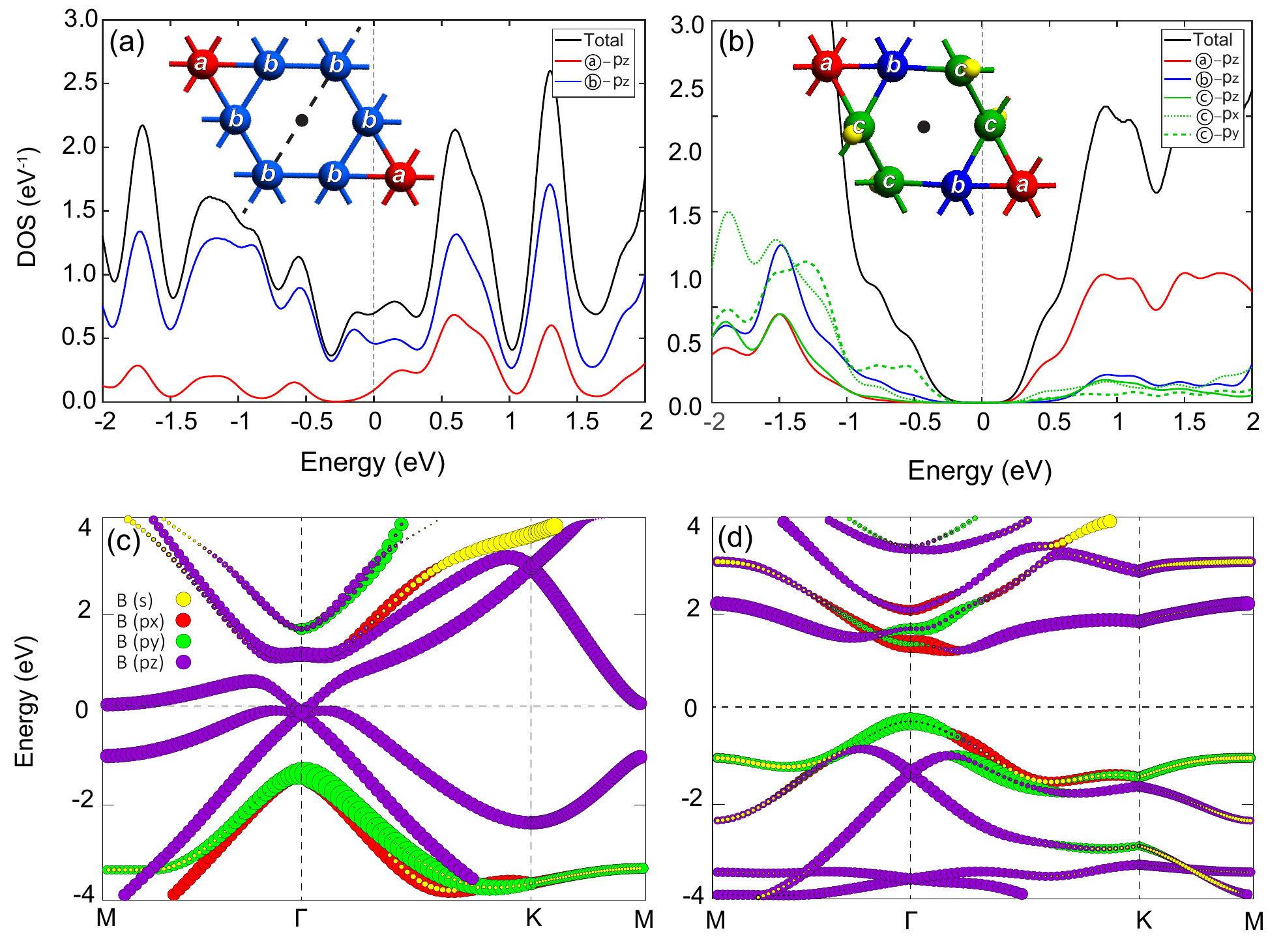}}
	\caption{\textbf{(a, b)} Partial density of states (PDOS) of $\alpha'$ and 
		\mbox{$\alpha'$-4H} borophenes, showing the significant contributing 
		orbitals 
		of 
		the labeled atomic sites shown in the insets. Full results of PDOS are 
		also 
		available in FIG.~S1. \textbf{(c, d)} Orbital decomposed band 
		structures of 
		$\alpha'$ and \mbox{$\alpha'$-4H} borophenes. Both PDOS and band 
		structure were 
		calculated at the GGA-PBE level of theory.}
	\label{fig2}
\end{figure*}

FIG.~\ref{fig2} summarizes the structural and electronic properties of 
$\alpha'$ and $\alpha'$-4H borophenes. As shown in the inset of FIG.~\ref{fig2} 
(a, b), the atomic sites in 
$\alpha'$ 
borophene 
are classified into 6- and 5-bonded boron atoms, labeled with \textcircled a 
and \textcircled{b}, 
respectively. There are two \textcircled{a} and six \textcircled{b} 
sites 
in each unit cell and we can see both inversion and mirror symmetry in 
$\alpha'$ borophene. After 
hydrogenation, four \textcircled{b} sites break a B-B bond and make a B-H 
bond instead, which leads 
to novel sites labeled with \textcircled{c}. As a result, hydrogenation breaks 
the mirror symmetry and only preserves the inversion symmetry in 
\mbox{$\alpha'$-4H} borophene. These 
changes 
directly affect the electronic states and should be considered as one of 
the origins of different properties in the two borophenes.
 
FIG.~\ref{fig2} also represents the partial density of states 
(PDOS) and orbital 
decomposed 
band structures of $\alpha'$ and \mbox{$\alpha'$-4H} borophenes at the GGA-PBE 
level. 
This information gives valuable insights into the contribution of each atomic 
site and 
orbital in 
the electronic properties. As it is clear, in $\alpha'$ borophene (left panel), 
$p_z$ orbitals of the \textcircled{b} sites are dominant. Given that $\alpha'$ 
borophene is 
almost flat, $p_z$ orbitals are perpendicular to neighboring $p_{x(y)}$ 
orbitals. 
Specifically, no hybridization takes place between these 
orbitals. In the band structure, we can see a triple energy degeneracy which is 
similar to a Dirac point, pierced by a third band. The triple band touching 
Dirac fermions was previously observed in other flat borophenes in experimental 
and theoretical studies \cite{ezawa17, feng17}.

In \mbox{$\alpha'$-4H} 
borophene (right 
panel), 
hydrogenation pulls some of \textcircled{b} atoms, labeled 
with \textcircled c, out of the plane and forms a buckled structure. As a 
result, $p_z$ orbitals 
of \textcircled{c} atoms would no longer 
be 
perpendicular 
to neighboring $p_{x(y)}$ orbitals, which leads to hybrid bonds. This 
hybridization 
can be seen very well in the PDOS and band structure. By hydrogenation, 
\textcircled{b} sites lose their dominance around the Fermi level. However, 
\textcircled{a} 
sites rather 
reserve their contribution, especially in the conduction band, because 
hydrogenation do not affect them directly. Subsequently, $\alpha'$-4H 
borophene has dominated 
occupied \textcircled{c}-$p_y$ states and unoccupied 
\textcircled{a}-$p_z$ states in the valance and 
conduction band edges, respectively, which are separated with a relatively wide 
band gap.

These results are completely supported by the shapes of the wave 
functions at the VBM and CBM as represented in FIG.~S2. At the VBM, the wave 
function is shaped like a dumbbell, which is distributed along the $y$-axis and 
centered 
on \textcircled c atoms, showing the contribution of \textcircled{c}-$p_y$ 
orbitals. At 
the CBM, the 
dumbbell-shaped wave function has components along both $x$-axis and $z$-axis, 
showing the 
hybridization between \textcircled{c}-$p_x$ and \textcircled{a}-$p_z$ 
orbitals.

To highlight the role of buckling and orbital hybridization as a consequence, 
we also calculated the band structure for a conceptual model of flat 
\mbox{$\alpha'$-4H} borophene. We re-optimized \mbox{$\alpha'$-4H} borophene 
with a fixed 
$z$-component, to have the same buckling as $\alpha'$ borophene. The 
resulting band structure is 
available in FIG.~S3 and shows no band gap. Therefore, we can conclude that 
the buckling induced by \textcircled{c} atoms is a strong cause for 
opening the band gap. It should be noted that the role of buckling in opening 
energy band gap has been widely considered in the community \cite{radha20, 
umeno19, 
nijamudheen15}. 

The band gap calculated at the GGA-PBE level is 1.49~eV, which is 
underestimated by 1.0~eV compared to the experiment \cite{hou20}. Using the 
HSE03 and 
HSE06 hybrid functionals and considering a fraction of the exact exchange, the 
band gap is corrected to be 1.85 and 1.98~eV, respectively (see FIG.~S4). 
Eventually, at the PBE0 level, the band gap is found to be 2.54~eV, in 
excellent agreement with the experimental band gap measured by UV-V spectrum 
(2.49~eV).

Although DFT gives adequate qualitative information about the electronic and 
optical properties, it does not account for the electron-electron and 
electron-hole interactions, which have a crucial role in 2D materials. 
Therefore, further investigations should be performed using many-body 
perturbation calculations. %, especially in terms of optical properties.

The QP band gap predicted by many-body GW theory is very sensitive to input 
parameters including the spin-orbit coupling (SOC), \mbox{$k$-point} sampling, 
vacuum 
space, number of virtual bands, energy cutoff, and self-consistent iteration. 
Therefore, it is mandatory to converge the band gap. We 
initially checked the effects of the SOC interaction. Due to small atomic mass 
of boron, except a slight downward 
shift in the electronic states, no noticeable spin splitting was found in the 
energy levels, and the band gap remained almost constant ($\Delta$E$_g$ 
$\simeq$ 
0.001 
eV).
Hence, due to its excessive computational costs and negligible impacts, this 
interaction is excluded from the rest of the calculations. 

Next, we performed a 
bunch of calculations to converge the QP band gap at the G$_0$W$_0$ level as a 
function of the (a) \mbox{$k$-point} sampling of Brillouin zone using 40 
virtual bands 
and \mbox{$L_z$ = 20~\AA}; (b) vacuum space using 40 virtual bands and \mbox{5 
$\times$ 5 
$\times$ 1} \mbox{$k$-point} 
mesh; and (c) the number of virtual bands using $L_z$ = 20~\AA\ and 5 $\times$ 
5 
$\times$ 1 
\mbox{$k$-point} 
mesh as illustrated in FIG.~S6. We realized that a 9 $\times$ 9 $\times$ 1 
\mbox{$k$-point} mesh is 
enough for a satisfactory convergence within $\sim$~0.05~eV. Such a dense 
\mbox{$k$-point} 
mesh is also indispensable for an accurate description of the optical gap and 
binding energy. The vacuum space plays a crucial role in the determination of 
the QP band gap 
due 
to the long-range nature of screened Coulomb interaction. Generally, it is 
imposed to be between 20 to 30~\AA. To ensure convergence within $\sim$~
0.05~eV, 
we 
set the vacuum space to be 25~\AA, which is the highest accuracy we could 
computationally afford. Our results also verified that at least 40 virtual 
bands are required for our many-body GW calculations. The corresponding band 
gap is essentially the same when 80 bands are included (see FIG.~S6). 
Such rapid convergence concerning the number of unoccupied bands is mainly 
attributed to the absence of strongly localized states characterized by flat 
bands. Moreover, we tested the convergence of the QP band gap with respect to 
the number of iterations in partially self-consistent G$_i$W$_0$ and G$_i$W$_i$ 
as 
shown in 
FIG.~S6. Obviously, four iterations are sufficient to reach the 
convergence threshold of $\sim$~0.001~eV. Also, the energy cutoff for the 
exchange 
and correlation parts of the self-energy was set to be 50 and 10 Ry, 
respectively.

To ensure the accuracy of our many-body GW calculations, we calculated the QP 
band gaps of graphane (CH) and fluorographene (CF) monolayers. For example, for 
graphane, the band gap was predicted to be 5.62, 5.81, and 6.17~eV under the 
G$_0$W$_0$, GW$_0$, and GW levels of theory, respectively, which are in 
excellent 
agreement with 5.64, 5.89, and 6.28~eV reported in previous works 
\cite{karlicky13}. 
Also, the band gaps obtained for fluorographene agree very well with previous 
values as listed in Table S2.

After validation of the required accuracy, now we turn our attention to the GW 
results. The general shape of the QP band structure at the G$_0$W$_0$ 
level is similar 
to 
that 
obtained by the PBE functional, shown in FIG.~S4. The VBM and CBM still locate 
at 
the same points. The Fermi level is still closer to the VBM, making 
\mbox{$\alpha'$-4H} 
borophene a p-type semiconductor. The valence and conduction band edges are 
symmetrically dispersed, which indicates the presence of free charge carriers. 
However, the inclusion of electron-electron interaction enlarges the band gap 
from 1.49 up to 1.98~eV. This value is larger than that of the HSE03 and 
exactly equal to that obtained by the HSE06. In other words, a similar accuracy 
could be achieved by the G$_0$W$_0$ as by the HSE06. At the GW$_0$ and GW 
levels, 
the 
QP 
band gap increases up to 2.23 and 2.52~eV, respectively, which is due to the 
presence of self-consistency. Overall, one can say that the GW level offers 
an accuracy similar to the PBE0 functional, in excellent agreement with the 
experiment \cite{hou20}. Also, the direct QP band gap at the $\Gamma$ point are 
3.09, 
3.31, 
and 3.58~eV for G$_0$W$_0$, GW$_0$, and GW, respectively.

\subsection{Optical properties}
\begin{figure}
	\includegraphics[width=0.45\textwidth]{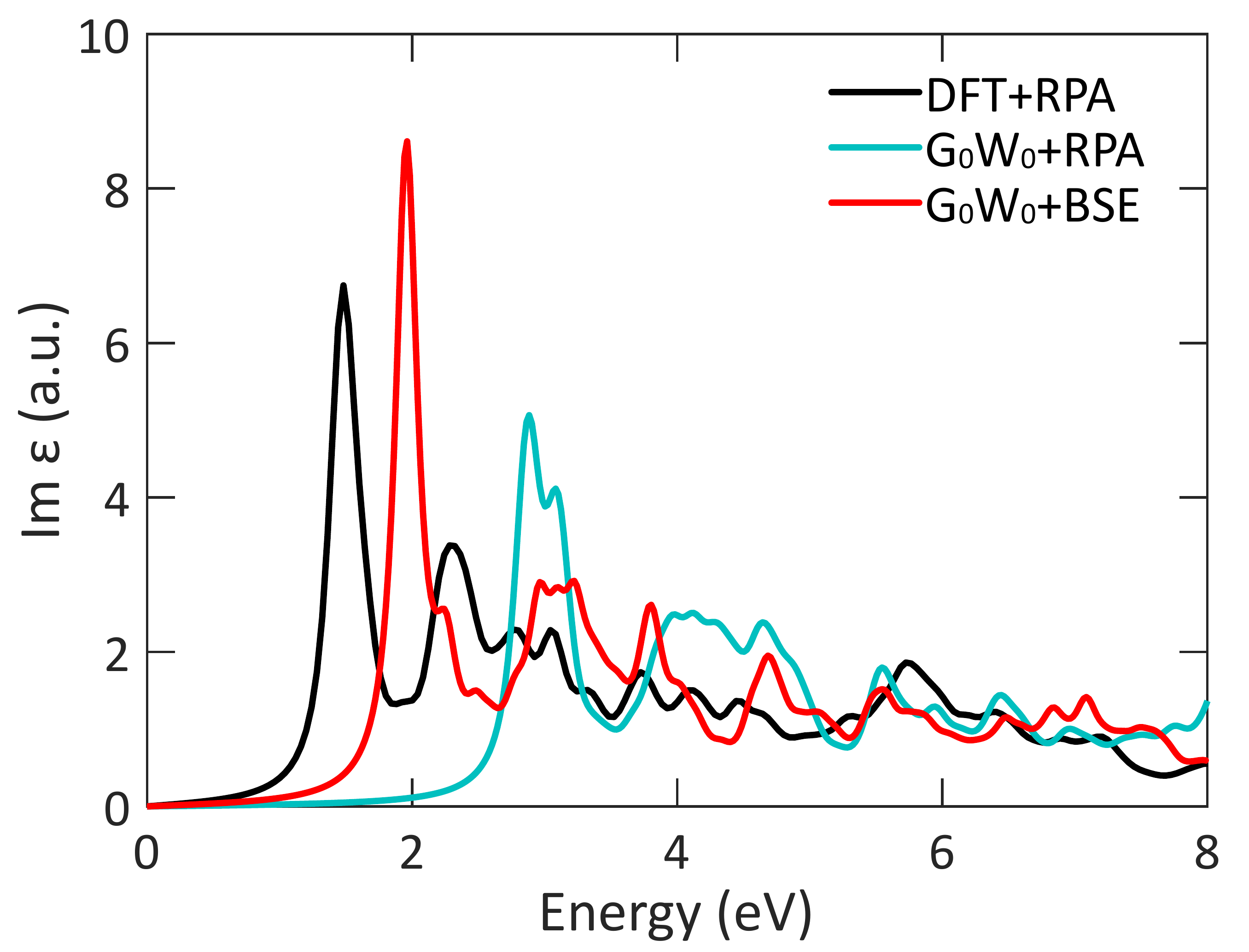}
	\caption{The imaginary part of the dielectric function of 
	\mbox{$\alpha'$-4H} 
	borophene for light polarized along the $x$-direction at different levels 
	of 
	theory: DFT+RPA (e-e and e-h correlations neglected), G$_0$W$_0$+RPA (e-e 
	correlation included and e-h correlation neglected), and G$_0$W$_0$+BSE 
	(both 
	e-e 
	and e-h correlations included).}
	\label{fig3}
\end{figure}

Excitonic effects play a crucial role in the optical properties of 2D materials 
for their weak dielectric screening. Hence, to have a reliable description of 
the optical properties, it is indispensable to take the electron-hole Coulomb 
interaction into account. In this section, we aim to present the excitonic 
optical spectra of \mbox{$\alpha'$-4H} borophene achieved from solving the 
Bethe-Salpeter equation on top of various levels of the GW. To fully 
appreciate 
the many-body effects, we also calculated the optical spectra within 
random-phase approximation (RPA) on top of DFT and GW levels of theory, where 
the electron-hole interaction is neglected. FIG.~\ref{fig3} exhibits the 
imaginary 
part of the macroscopic dielectric function for light polarized along the 
$x$-direction at different levels, namely DFT+RPA, G$_0$W$_0$+RPA, and 
G$_0$W$_0$+BSE. As can 
be seen, the inclusion of electron-electron correlation completely reshapes the 
optical spectrum and leads to a significant blue shift ($\sim$~1.4~eV), which 
is due 
to the self-energy correction. On the other hand, the electron-hole correlation 
results in a 
redshift ($\sim$~0.92~eV) in the optical spectrum, owing to the cancellation 
effect 
between self-energy correction and excitonic effects. The electron-hole 
correlation also leads to an increase (decrease) of the oscillator strength 
of 
the first 
(second) peak. However, its main impact is the appearance of some bound 
excitons below the G$_0$W$_0$ direct band gap (3.09~eV) at the $\Gamma$ point, 
which 
are 
missing in the G$_0$W$_0$+RPA spectrum. As the imaginary part of the dielectric 
function is directly associated with the absorption spectrum, one can say that 
the excitonic effects increase the light absorption in the visible range (1.5 
to 3.5~eV).

	\begin{figure}
	\includegraphics[width=0.45\textwidth]{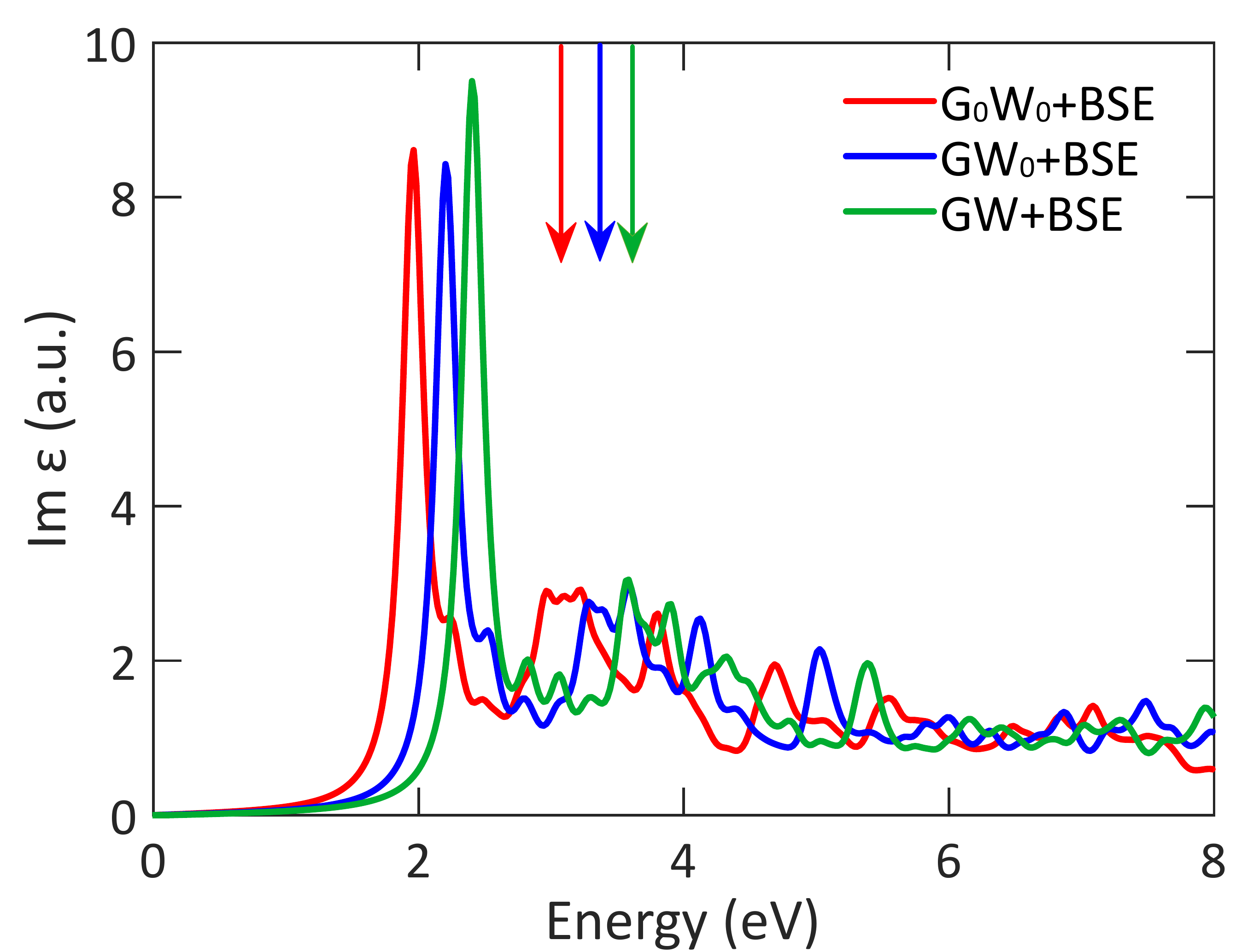}
	\caption{The imaginary part of the dielectric function of 
	\mbox{$\alpha'$-4H} 
	borophene including excitonic effects for light polarized along the 
	$x$-direction under different degrees of self-consistency in the GW. 
	Four 
	iterations 
	were 
	used to 
	obtain well-converged results from G$_i$W$_0$ and G$_i$W$_i$ calculations. 
	The QP (direct) band gaps are shown with arrows.}
	\label{fig4}
\end{figure}

The DFT+RPA predicts the absorption edge to be $\sim$~1.48~eV, which is very 
close 
to the PBE band gap (1.49~eV). The first two peaks at G$_0$W$_0$+RPA appear at 
2.88 
and 3.08~eV, respectively. They are near the G$_0$W$_0$ direct band gap (3.09 
eV) 
at 
the $\Gamma$ point, which is due to the exclusion of excitonic effects. Most 
importantly, the first peak at G$_0$W$_0$+BSE ($\sim$~1.96~eV) is found to lie 
between the 
PBE (1.49~eV) and G$_0$W$_0$ (1.98~eV) band gaps as it includes both the QP 
correction and excitonic effects. This peak (the so-called optical gap) 
originates from the direct transition (bright exciton) from the VBM to the 
conduction band, and corresponds to a strongly bound Frenkel exciton with a 
binding energy of $E_b$ $\simeq$ 1.13~eV. It is worth mentioning that the 
exciton 
binding 
energy is determined by the difference between the QP direct band gap and 
optical gap. The optical gap obtained from G$_0$W$_0$+BSE is $\sim$~0.53~eV 
smaller 
than 
the experimental gap achieved from photoluminescence measurement (2.49~eV) 
\cite{hou20}. This underestimation is attributed to the overscreening effects 
associated with building the susceptibility based on an underestimated 
Kohn-Sham band gap. Indeed, the small Kohn-Sham band gap leads to a spurious 
enhancement of the screening, and subsequently, to an underestimated optical 
gap.

For the considerable dependence of the G$_0$W$_0$ on the Kohn-Sham starting 
point 
and 
the exclusion of many-body exchange and correlation in the ground-state 
eigenvalues, we examined the effects of self-consistency. FIG.~\ref{fig4} 
indicates 
the optical spectra of \mbox{$\alpha'$-4H} borophene for light polarized along 
the 
$x$-direction under different degrees of self-consistency in the GW. As can 
be 
seen, iteratively updating the QP eigenvalues in the Green’s function results 
in a blue shift ($\sim$~0.24~eV) in the optical spectrum because it reduces the 
overestimation of the screening observed in the G$_0$W$_0$. At this level of 
theory 
(i.e. GW$_0$+BSE), the shape of the spectrum is relatively preserved, and the 
optical gap is estimated to be $\sim$~2.20~eV with an exciton binding energy of 
$E_b$ $\simeq$~
1.11~eV. Therefore, one can say that $E_b$ is independent of the 
self-consistency 
in Green’s function. Moreover, the oscillator strength is only redistributed to 
higher energies with no considerable change.

Updating the screened Coulomb potential leads to a blue shift ($\sim$~0.20~eV) 
as 
well, implying a larger self-energy correction in the GW compared to those in 
the G$_0$W$_0$ and GW$_0$. The highest level of theory (i.e. GW+BSE) predicts 
the 
optical gap to be $\sim$~2.40~eV, which is slightly smaller than the GW band 
gap of 
$\sim$~2.52~eV. This value agrees very well with the experimental gap (2.49~eV) 
\cite{hou20} 
because GW+BSE incorporates many-body effects in the ground state eigenvalues 
and meaningfully reduces the dependence on the DFT starting point using 
self-consistency. Here, the $E_b$ is obtained $\sim$~1.18~eV, suggesting 
relatively 
stronger cancellation effects. Such a huge $E_b$ is a signature of a strongly 
bound Frenkel exciton, showing ultrahigh stability of excitonic states against 
thermal dissociation, making \mbox{$\alpha'$-4H} borophene a very promising 
candidate 
for exploring room-temperature optical applications. The final value is larger 
than those of phosphorene (0.86~eV), MoS$_2$ (0.96~eV), 
h-BN 
(0.80~eV), and many other semiconducting monolayers \cite{qiu13, kolos19, xu17, 
tran14}, turning 
\mbox{$\alpha'$-4H} borophene into a highly competitive candidate. As the 
GW+BSE gives 
the best results, the corresponding oscillator strength is shown in FIG.~S7.

Because of the huge depolarization effect in 2D systems for the light 
polarization perpendicular to the surface (E$\Vert z$) \cite{yang07}, we only 
focus on 
the 
optical spectra for light polarization parallel to the surface (E$\Vert x$, 
E$\Vert y$). As 
a consequence of the structural anisotropy in \mbox{$\alpha'$-4H} borophene, a 
different optical spectrum is obtained for light polarized along the 
$y$-direction. As shown in FIG.~S8, at all three levels of theory, the shape of 
the spectrum is different from that for light polarized along the $x$-direction 
(FIG.~\ref{fig4}). Also, the first peak has weaker intensity. 
However, 
the same optical gap is achieved. Such in-plane optical anisotropy provides an 
additional degree of freedom for manufacturing optoelectronic devices.

It is worth mentioning that the QP band gap and exciton binding energy 
calculated at all three levels of theory (i.e. G$_0$W$_0$, GW$_0$, and GW) 
show 
an 
excellent agreement with the linear scaling law given as, 
$E_b$~=~0.21$~E_g$~+~0.40 
\cite{choi15}. The standard deviations are less than 0.05~eV, signifying 
well-converged 
results.

We also investigated the variation of band gap under biaxial tensile strain. As 
can be seen from FIG.~S9, by increasing the tensile strain up to 10\%, the QP 
and optical band gaps decrease linearly by $\sim$~0.4 and 0.7~eV, respectively. 
This 
is due to the reduction of buckling under tension. The exciton binding 
energy also experiences a decrease of $\sim$~0.3~eV, which means the excitonic 
states 
are robust against tension up to 10\%. It should be noted, in our previous 
work, we 
had presented the dynamical stability of $\alpha'$-4H borophene under such 
tension by phonon dispersion, stress-strain, and energy-strain curves
\cite{mozvashi21}.

%Such tunable optical band gap and 
%large exciton binding energy suggest \mbox{$\alpha'$-4H} borophene as a 
%promising candidate for optical applications at room temperature.

\subsection{Bilayer \mbox{$\alpha'$-4H} borophene}
\begin{figure*}[t]
	\includegraphics[width=0.8\textwidth]{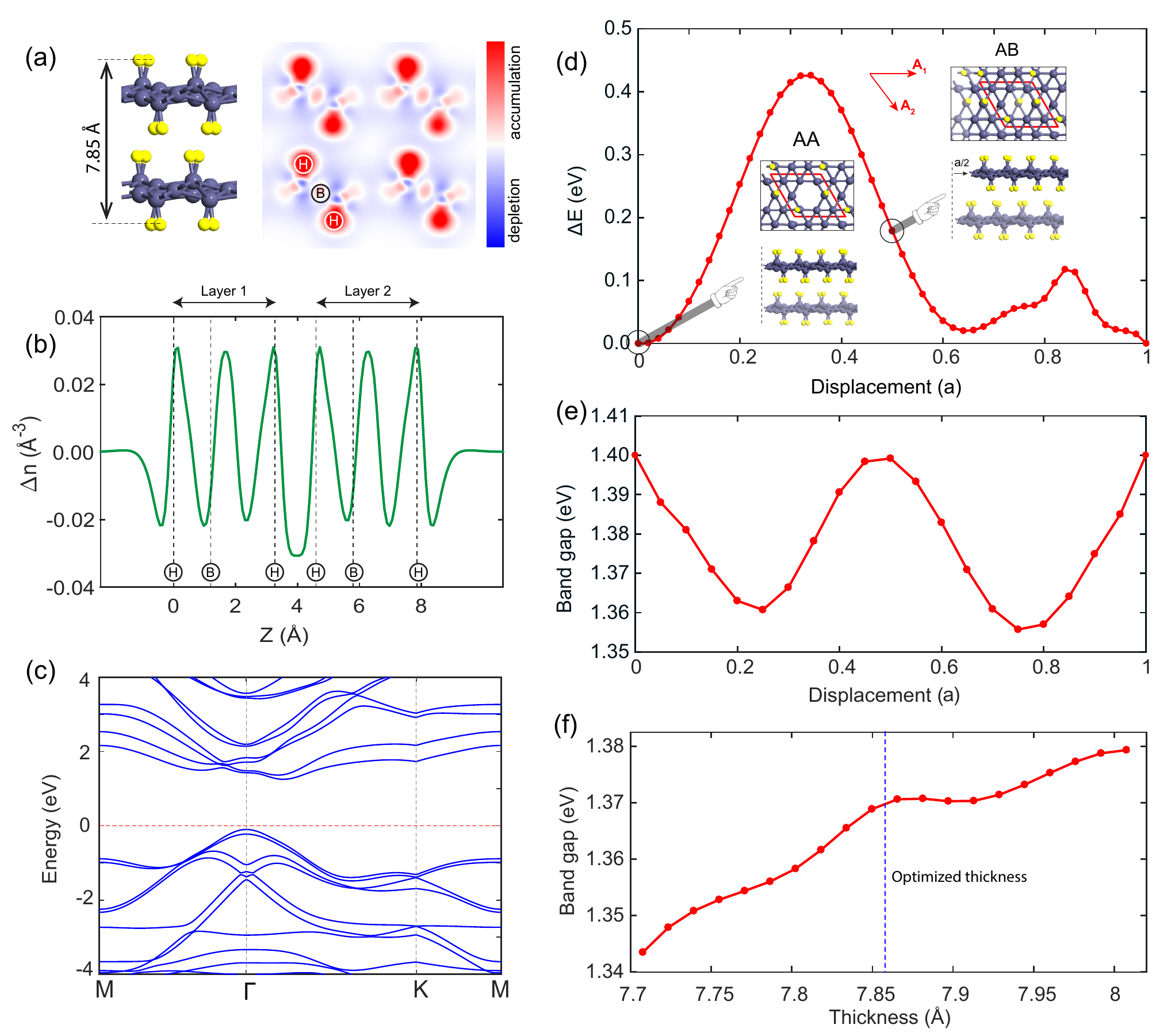}
	\caption{Bilayer \mbox{$\alpha'$-4H} borophene. \textbf{(a)} Side view and 
	electron 
	difference 
	density map. \textbf{(b)} Variation of electron difference density with 
	buckling. 
	\textbf{(c)} Band structure of AA stacking. \textbf{(d)} Variation of 
	bilayer energy 
	with 
	displacement of the upper layer with respect to AA stacking. \textbf{(e)} 
	Variation of bilayer band gap with displacement of the 
	upper 
	layer. \textbf{(f)} Variation of band gap of AA stacking with thickness.}
	\label{fig5}
\end{figure*}

Observations of atomic force microscopy (AFM) have shown that the average 
thickness of a typical synthesized \mbox{$\alpha'$-4H} borophene is about 0.78 
nm 
\cite{hou20}, which is higher than the monolayer. To understand the effects of 
thickness on 
the electronic properties of \mbox{$\alpha'$-4H} borophene, we studied the 
bilayer of 
this material. For this, we firstly measured the vertical distance between 
layers by means of variation of energy to thickness. As FIG.~\ref{fig5} (a) 
shows, the 
optimized thickness of a bilayer is achieved 0.785 nm, which is in 
great agreement with measurements of AFM. After gaining the 
interlayer distance, we re-optimized the structure with AA stacking. The 
electron difference density in FIG.~\ref{fig5} (a \& b) shows that while a high 
electron extracting takes place by hydrogen atoms, the two layers do 
not trade 
much 
electrons within themselves, which implies the absence of strong covalent 
bonds and presence of weak van der Waals interactions. This might be due to 
the screening of H atoms among the boron layers. 
Because of the weak force between the layers, the 
lattice constant and bond lengths of the bilayer do not have a significant 
difference with the monolayer.

As shown in FIG.~\ref{fig5} (c), the AA-stacked bilayer \mbox{$\alpha'$-4H}  
borophene
is an 
indirect semiconductor with a band gap of 1.4~eV. Most 2D materials experience 
a topological band gap transition from their monolayer to bilayer or multilayer 
forms. For instance, from monolayer to bilayer, graphene opens a bang gap of 
zero to 0.25~eV \cite{zhang09} and, on the contrary, $\beta$ antimonene closes 
its band 
gap 
from 0.76~eV to zero \cite{wang15}. However, for the weak van der Waals 
interlayer 
interaction in bilayer \mbox{$\alpha'$-4H} borophene, it experiences a very 
small gap 
change. The VBM is located at $\Gamma$ point, and the CBM is within the 
$\Gamma$-K path. The 
bilayer \mbox{$\alpha'$-4H} borophene is a p-type semiconductor because VBM is 
very 
close to the Fermi energy. This is due to electron deficiency in boron-based 
structures. %Also, the asymmetry in band extrema was previously reported in C3N 
%monolayers [56]. 
In the VBM and CBM, the band structure dispersion is 
parabolic, which exhibits high mobility of electrons and holes. This inclines 
the high potential of \mbox{$\alpha'$-4H} borophene in optoelectronic 
applications,~eVen for thicknesses more than monolayer.

To see the impact of stacking mode on the electronic properties of bilayer, we 
slipped the upper layer along the lattice vector \textbf{A}$_1$
as much as 
percentages of lattice constant, a. The energy difference between different 
stackings with respect to AA stacking is shown in FIG.~\ref{fig5} (d), which is 
less than 
20 
meV. These stackings can form Moiré patterns in borophene with no need 
to 
external 
strain or requirement of large superlattices. Our investigations show that 
stacking mode does not have significant impact on the band gap of borophene. 
The 
band gap remains indirect in all the stackings and, as shown in FIG.~
\ref{fig5} 
(e), it fluctuate mildly around the band gap 
of the AA stacking. When the displacement changes as half of 
the 
lattice 
constant (AB stacking), the band gap becomes equal to the band gap of the AA 
stacking. Because different stackings do not have very different band gaps, it 
is 
not 
possible to determine the stacking mode of bilayer $\alpha'$-4H borophene 
through absorption 
spectrum.

To understand the dependence of electronic properties on the vertical distance 
between the layers, variation of band gap with interlayer distance is shown in 
FIG.~\ref{fig5} (f). Herein, with conserving the atomic arrangements, we tuned 
the interlayer distance to have thicknesses between 7.7 to 8~\AA. 
We can see a mild 
increase of band gap with the increase of the thickness. In other words, a high 
increase in the distance cancels the weak van der Waals interaction between 
layers and renders the bilayer into two independent monolayers. On the 
contrary, 
decrease of vertical interlayer distance increases the interactions and leads 
to a mild decrease in the band gap. Our investigations show that interlayer 
distance does not have any impact on the nature of indirect band gap of 
\mbox{$\alpha'$-4H} borophene. It can be concluded that for the weak van der 
Waals interaction between the layers and existence of light weight atoms, 
vertical 
pressure and slipping of the layers do not make 
significant change in optical properties of the structure, therefore, no gap 
difference was observed for different thickness of the structure in experiment 
\cite{hou20}.
 
\section{Conclusion}

In summary, we reported the electronic and excitonic optical properties of 
$\alpha'$-4H borophene using first-principles many-body calculations. We 
firstly tracked the mechanism in which the metallic $\alpha'$ borophene turns 
into 
the semiconducting $\alpha'$-4H borophene. It was found that hydrogenation 
increases the buckling height and breaks the mirror 
symmetry of $\alpha'$ borophene, which results in 
hybridization between $p_z$ and neighboring $p_{x(y)}$ orbitals and opens a 
wide band gap in $\alpha'$-4H borophene. Then, we calculated the 
quasiparticle 
bang gap to be 1.98, 2.23, and 2.52~eV at the 
G$_0$W$_0$, GW$_0$, and GW levels of theory, respectively. In the following, we 
solved the 
Bethe-Salpeter equation on top of the GW. The optical spectrum at GW+BSE shows 
that the direct optical excitation is characterized by an optical band gap of 
$\sim$~2.40~eV, which is in excellent agreement with the experiment (2.49~eV). 
The exciton binding energy was calculated to be $\sim$~1.18~eV, which is 
a signature of a strongly bound Frenkel exciton, suggesting ultrahigh stability 
of excitonic states against thermal dissociation. We also investigated the 
bilayer $\alpha'$-4H borophenes with different stackings and thicknesses. If 
was 
found that they have similar electronic properties to the monolayer, which 
gives them similar potentials.
These features together with great air-stability and mechanical strength 
suggest \mbox{$\alpha'$-4H} borophene as a desirable material for 
flexible optoelectronic 
applications at room temperature.

\section*{Acknowledgment}

We are thankful to the Research Council of the University of Guilan for the
partial support of this research.

\section*{Declaration of Interests}
The authors declare that they have no conflict of interest.

\bibliographystyle{apsrev4-2} % Tell bibtex which bibliography style to use
\bibliography{Ref}
\end{document}